\documentstyle[prl,aps,preprint]{revtex}
\begin{document}
\draft
%\twocolumn
\title{Adiabatic mechanism of the multiply charged ion
production by a laser field through ATI states of an atom}
\author{M.~Yu.~Kuchiev\cite{byline}}
\address{School of Physics, University of New South Wales,
Sydney 2052, Australia}
\date{\today}
\maketitle

\begin{abstract}
ATI can be followed by
an inelastic collision of the ionized electron with the parent atomic
particle resulting in an excitation of the ion.
It may be a continuum state excitation producing 
the doubly charged ion
or a discrete state which also 
enhances the doubly charged ion production. 
Absorption of a few quanta above the atomic threshold is sufficient to make
this mechanism work. As a result
the two-electron processes can take place even in  moderate fields.
The example of two-electron excitations of He atoms 
in a 780 nm laser field with intensity above $ 10^{14}$W/cm$^2$
is discussed.
\end{abstract}

\vspace{10mm}
PACS:3280k

%\pacs{PACS:3280k}

\newpage

The purpose of this paper is to demonstrate that ATI plays a very
important role in  multiply charged ion formation in a laser
field. 
The basic physical idea is as follows.
Suppose that   single electron ionization takes place. After that
the ionized electron 
can undergo an inelastic collision with the parent atomic particle.
The electron impact can result in the
excitation of the  ion into a discrete or continuum state because
the ionized electron, strongly 
interacting with the laser field  absorbs and  accumulates  the high energy
necessary for the ion excitation. 
This energy manifests itself in the form of the
 electron wiggling in the laser field as well as in the translational motion
of the electron. Consider the classical 
momentum of the electron. The laser field results in the oscillation
of the momentum ${\bf p}(t)$,
${\bf p}(t)={\bf k}+{\bf f}\sin \omega t$. Here the first constant term
${\bf k}$ describes the translational motion  and the second term,
${\bf f} \sin \omega t$, where ${\bf f}={\bf F}/\omega$,
describes the wiggling of the electron 
in the laser field which is considered to be linearly
polarized: ${\bf F}(t)={\bf F} \cos \omega t$.
As a result the  ``energy'' $E_{\bf k}(t)$ of the electron is also 
time-dependent, 
it behaves like 
\begin{equation} \label{Ek}
E_{\bf k}(t)=\frac{1}{2}\left ({\bf k}+
{\bf f} \sin \omega t \right )^2.
\end{equation}
For a given level in the ATI spectrum 
the translational momentum  
 satisfies the energy conservation law
$k^2/2+f^2/4+E^{(A^+)}=n_1\omega +E^{(A)}$, where $E^{(A)}$ and $E^{(A^+)}$
are the energies of the atomic and ionic ground states, $n_1$ is the number
of absorbed quanta during the single electron ionization.
Obviously for higher energy levels 
in the ATI spectrum, $i.e.$ for  larger  $n_1$, 
  the translational momentum is greater.
(For  description of ATI see 
Refs.\cite{ago,bor}
%(Agostini $et~al$ 1979), 
%(Boreham and Hora 1979) 
and  references in the review
\cite{fre}.)
%(Freeman and Bucksbaum 1991).)
We will show below that 
double ionization caused by  inelastic scattering of the firstly
ionized electron on the parent atomic particle is possible
if the energy necessary for removal of the second electron satisfies a 
condition 
\begin{equation}\label{maxE}
E^{(A^{++})}-E^{(A^{+})}
\le 
\frac{1}{2}\left ( k + f \right ) ^2~,
\end{equation}
where $E^{(A^{+})}$ and $E^{(A^{++})}$ are the ground states of the single
and doubly charged ions.
Eq.(\ref{maxE})
has a clear  physical meaning. Consider a given energy level in the  ATI 
spectrum. Then the right-hand side of Eq.(\ref{maxE}) is equal to the 
maximum
of $E_{\bf k}(t)$ (\ref{Ek}), ${\rm max}[E_{\bf k}(t)]=(1/2)(k+f)^2$, 
which is achieved when  ${\bf k}/k=\pm {\bf F}/F$ and $\omega t=\pm \pi/2$. 
Inequality (\ref{maxE}) simply
states that the maximal 
energy of the photoelectron, which is possible for a given $n_1$, 
must exceed the 
ionization energy.
If the field is strong enough, $i.e.$
$f^2/2\ge E^{(A^{++})}-E^{(A^{+})}$, 
then according to Eq.(\ref{maxE}) it is sufficient to
consider the lowest level in the ATI spectrum for which $k\approx 0$. 
For weaker fields the translational momentum 
becomes vital and therefore ATI comes into play.
For a sufficiently high level in the ATI spectrum  
the translational momentum becomes high enough to satisfy
the  inequality in Eq.(\ref{maxE}) permitting the mechanism to work.
Note that there is 
no sharp threshold for the field intensity, below which the mechanism fails.
It always works, but the weaker the field, the higher the necessary 
level in the ATI spectrum.

It is important that according to Eq.(\ref{maxE})  
the necessary level in the ATI spectrum 
can be well below the energy necessary for removal
of the second electron. The above-threshold
energy $E_{\rm ati}$
of the level in the ATI spectrum $E_{\rm ati}=k^2/2+f^2/4$
is always lower than the maximal kinetic energy of the photoelectron
on this level, 
$k^2/2+f^2/4<{\rm max}[E_{\bf k}(t)]=(1/2)(k+f)^2$. 
The stronger the field, the more pronounced  the 
difference between 
$E_{\rm ati}$ and ${\rm max}[E_{\bf k}(t)]$. As a result
${\rm max}[E_{\bf k}(t)]$ can exceed the ion excitation energy while
$E_{\rm ati}$ can be well below it, 
$E_{\rm ati}<E_{\rm exc}<{\rm max}[E_{\bf k}(t)]$.
This means that absorption
of a small number of quanta above the atomic threshold 
can put the considered mechanism in action, resulting in  the population
of the doubly charged ion state.

The firstly ionized electron, which plays a crucial role in the considered 
mechanism, can be looked at as a kind of  antenna. It accumulates
the energy from the field and transfers it to the parent atomic particle.
This idea was first considered in 
Ref.\cite{k1}.
%(Kuchiev 1987). 
It was shown in this work that the scattering 
of the ionized electron on  the parent ion
can result in the ion excitation
or double ionization. It can  also  increase  the energy of the 
firstly
ionized electron populating the very high levels in the ATI spectrum. 
These results were based on the
analytical calculation of the amplitude
of the double ionization process. The calculations are  vital 
to justify the physical idea of rescattering. 
Without them
 one can doubt that a collision of the ionized electron with
the ion is possible at all because during single electron ionization
the ionized electron can go far outside the atom and never return. 
The calculations described in 
Ref.\cite{k1}
%(Kuchiev 1987) 
and
presented in detail in 
Ref.\cite{k2}
%(Kuchiev 1995) 
show that there is no danger of this 
kind, the collision is very probable.
The physical idea of rescattering of the ionized electron on
the parent atomic particle was also discussed in the recent paper 
Ref.\cite{cor}.
%(Corcum 1993).
In this work a model approach was developed. Recent experiments 
with two color lasers on ATI spectra Ref.\cite{sch}
support the rescattering mechanism.
In Ref.\cite{wal} 
the results of precise measurements of double ionization
of Helium are  presented and compared with the rescattering mechanism. 
A conclusion of this work
is that the experimentally observed yield
of doubly charged ions is higher than predicted by
the rescattering mechanism. The recent work Ref.\cite{ku}
also indicates that the production of doubly charged 
due to the rescattering mechanism is smaller than the one observed
experimentally.
Note, however, that this discrepancy is not strong, one or two 
orders of magnitude. At the same time
there are uncertainties in the calculations.
They come from  uncertainties in the cross sections of
He$^+$ ionization and excitation in the laser field and 
uncertainties connected with the classical models developed
in Refs.\cite{cor} and \cite{ku}. 
Therefore, the rescattering mechanism  qualitatively agrees with
the recent experiments 
Ref.\cite{wal} as well as  with earlier experiments on noble gases
Refs. \cite{lhu1,lhu2,lhu3,lom,luk,boy,chi}.
%(L'Huillier $et~al$ 1982,1983a,1983b, Lompre $et~al$ 1984,
%Luk $et~al$ 1983, Boyer $et~al$ 1984, Chin $et~al$ 1985).

Let us demonstrate the validity of inequality (\ref{maxE}).
Consider an atom placed in the laser field
whose frequency  is much lower
than the excitation energy of the atom. 
The fact that the frequency is low permits one to consider the process as 
adiabatic.
This is the basis of the Keldysh solution
of the  single electron ionization problem, Refs.\cite{kel},\cite{rei}.
%(Keldysh 1964), (Reiss 1990).
Consider what happens with the system
after the single electron ionization takes place.
The adiabaticity allows one to describe the 
strong influence of the laser field on the electron system
using the time-dependent ``energy'' of the electron system.
Following the tradition of  the theory of adiabatic processes we will call 
this ``energy''  an electron $term$.
To simplify consideration we will neglect
the static Coulomb interaction between the ionized electron and the ion.
This approach was first used in 
Ref.\cite{kel}.
%(Keldysh 1964). 
It is supposed to give  a  reasonable
description of atomic ionization and  must be even better
for electron detachment from  negative ions. The system under consideration
is the single-charged ion and the ionized electron. 
We will neglect the influence 
of the laser field on the energy  $E^{({\rm }A^{+})}$
of the ion ground state because it is 
suppressed due to a high ionization potential of the ion.
In contrast, the energy of
the ionized electron exhibits strong variation. Eq.(\ref{Ek}) describing 
this
variation was evaluated from the classical point of view,
but it remains valid for 
the quantum description as well. This follows from the Volkov solution, 
Ref.\cite{vol},
for the electron wave-function in a laser field: $\psi_{\bf k}({\bf r},t)=
\exp \{ i[( {\bf k}+{\bf f}\sin \omega t){\bf r}-\int^t_0 E_{\bf k}(\tau) 
d\tau ]\}$, where 
$E_{\bf k}(t)$
in the exponent describes the term variation given in Eq.(\ref{Ek}).
Thus 
the term
$E^{({\rm e + A}^{+})}(t)$ of the system consisting of the single-charged 
ion 
and the photoelectron is
\begin{equation}\label{eA+}
E^{({\rm e + A}^{+})}(t)= E^{({\rm }A^{+})}+E_{\bf k}(t)~.
\end{equation}
Consider now the final state of the reaction where there are the doubly 
charged
ion and two electrons in the continuum. 
Let us neglect the static Coulomb field of the ion
as well as the Coulomb repulsion between the ionized electrons. 
This approximation is certainly more questionable than the similar
one for the single-electron ionization. Nevertheless, 
one can suppose  it to
be reasonable for double electron detachment from negative ions. 
It is argued below that the case of atomic ionization 
can also be approached using this approximation.
Considering  the energy of the doubly charged ion as a 
constant $E^{({\rm A}^{++})}$ and taking into account that the energies of 
both ionized electrons
oscillate in accordance with Eq.(\ref{Ek}) 
we find the final state term of the system consisting of the doubly
charged ion and the two ionized electrons
\begin{eqnarray}\label{EA++}
E^{({\rm e_1+e_2 + A}^{++})}(t)= E^{({\rm }A^{++})}+E_{{\bf k}_1}(t)+
E_{{\bf k}_2}(t)~.
\end{eqnarray}
Here ${\bf k}_1,~{\bf k}_2$ are the translational momenta of the two 
electrons in the continuum. They depend upon the total number $n$ of 
absorbed 
quanta $[k^2_1+f^2/4]+[k^2_2+f^2/4]+E^{(A^{++})}=
n\omega +E^{(A)}$, where $E^{(A)}$ is the atomic energy and
$n\omega$ exceeds the ionization potential, $n\omega
>E^{(A^{++})} -E^{(A)}+f^2/2$.

According to the adiabatic theory a transition from the term of the system
``electron + single charged ion'' 
to the  term ``two electrons + doubly charged ion''
can take place with high probability if 
there is a crossing of the two terms for some real moment of time $t$: 
\begin{equation}\label{cros}
 E^{({\rm e + A}^{+})}(t) = E^{({\rm e_1+e_2 + A}^{++})}(t)~.
\end{equation}
We are interested to find this crossing when the firstly ionized electron
occupies some low-lying level in the ATI spectrum because 
population of lower levels is the most probable.
The behavior of the two terms in Eq.(\ref{cros}) is illustrated in Fig.1
for different levels in the ATI spectrum. 
It is important to consider the final state term of the doubly charged ion
given in Eq.(\ref{EA++})
when both ionized electrons occupy  levels in the ATI spectrum,
above the ionization potential 
of the doubly charged ion. Then this term can possess minima at the moments
of time when the term of the single-charged ion Eq.(\ref{eA+})
possesses maxima.
 The example presented in
Fig.1 illustrates that the desirable crossing of the two
terms 
can be obtained if a few quanta are absorbed above
the atomic threshold, see the crossing of the term {\bf b}
with the terms {\bf d} and {\bf e}.

It is easy to verify that the crossing of terms in 
Eq.(\ref{cros}) is possible if inequality (\ref{maxE}) 
is fulfilled. Therefore, if (\ref{maxE}) is valid then
the transition from one term to the other is allowed
from the adiabatic point of view. 
In order to prove that  the transition really
takes place it is necessary to consider a potential 
having matrix elements mixing the two terms. 
The role of such a  potential is played by the interaction
between the  ionized electron and the single-charged ion.
It certainly has the necessary matrix elements: they describe the
ionization of the second electron caused by the inelastic scattering of the 
electron on the ion. 
Having the crossing of the two terms and the mixing potential one can 
calculate
the probability of the transition from one term to the other
using  perturbation theory. Technically this is a 
complicated problem because the electron in the system $e + A^+$
belongs to the continuum spectrum.
Nevertheless the qualitative final answer to this problem is simple 
Ref.\cite{k2}.
%(Kuchiev 1995).
The probability $W^{(2e)}$ of the double ionization is 
$W^{(2e)}\approx W^{(e)}\sigma_{\rm in}/R^2$. 
Here $W^{(e)}$ is the probability
of the population of the lowest level in the ATI spectrum
of the single-charged ion
which satisfies Eq.(\ref{maxE}), $\sigma$ is the inelastic cross section
of the electron impact on the single-charged ion in the laser field.
The quantity 
\begin{equation}\label{R}
R=\frac{F}{\omega^2}
\{ [(\beta+1)^2+\gamma^2][(\beta-1)^2+\gamma^2]\} ^{1/2},
\end{equation}
is a result of calculations. It
can be considered as an effective radius of the region 
in which the firstly ionized electron is localized.
It is proportional to the magnitude $F/\omega^2$ of the electron wiggling 
in the laser field and depends on  the Keldysh
adiabatic parameter $\gamma=[2(E^{(A^{+})}-E^{(A)})]^{1/2}/f$, and  
the  parameter $\beta=k/f$ describing 
the considered ATI state of the single-charged ion. 

%\newpage

%\begin{figure}[h]
%\input psfig
%The body of the figure goes here.
%\psfig{file=fig.ps,height=10cm,angle=0}

The considered inelastic collision of the ionized 
electron with the ion takes place when the absolute value of the
electron momentum is maximal. Therefore, the neglected
influence of the static Coulomb ion field on this electron 
should not be strong. The final-state electrons appear
with the small momenta, and one can expect that the
neglected static Coulomb field of the ion can  only increase the
probability of this event.
Note that we also neglected  the Coulomb repulsion between the 
final-state electrons. 
This was possible because the momenta of the two
electrons in the final state can be different, as illustrated in Fig.1
by curves {\bf e},{\bf d}.
These arguments are certainly not decisive and the role
of the Coulomb interaction in the final state
should be studied in more detail. 
We will not do it in this work,
but consider instead a related problem, in which
the final-state Coulomb interaction  is less important while the final
answer for the problem is  similar to the one considered above.

Consider  the 
single electron ionization with excitation of the ion into a discreet
state. In this case we have the excited single-charged ion
in the final state of the reaction.
This excited state
can be substantially influenced  by the laser field. In the 
simplest case, when the second
order of perturbation theory is valid, the shift of the excited level caused
by the field is $\Delta E^{\rm (A^{+*})}=-\alpha (\omega ) F^2/2$, where
$\alpha (\omega )$ is the dynamical polarizability of the considered
excited level which is assumed positive, $\alpha (\omega )>0$.
 In this approximation one finds the term of the system
consisting of the ionized electron and the excited ion 
\begin{eqnarray}\label{exte}
E^{\rm (e + A^{+*})}(t)=E^{\rm (A^{+*})}- 
\frac{1}{2}\alpha (\omega ) F^2 \cos^2\omega t +E_{\bf k}(t)~,
\end{eqnarray}
where $E^{\rm (A^{+*})}$ is the non-shifted position of the excited ion 
level.
Now we can consider the transition from the term describing the intermediate
state  $e + A^{+}$ to the term describing the final state $e+A^{(+*)}$.
This transition is possible if equation $E^{\rm (e + A^{+})}(t)=
E^{\rm (e + A^{+*})}(t)$ is resolved for a real moment of time $t$.
Using Eqs.(\ref{eA+}),(\ref{exte}) we find that the solution exists if
\begin{equation}\label{exc}
E^{\rm (A^{+*})}-E^{\rm (A^{+})}\le \frac{1}{2}\left ( k+f
\right )^2~.
\end{equation}
We see that the ion excitation is allowed if the necessary energy of the 
excitation is lower than the maximal energy of the photoelectron
on the considered level of the ATI spectrum. The resemblance between
Eqs.(\ref{exc}) and (\ref{maxE}) is obvious.

Notice that there is no sharp energy cut-off for the considered mechanism. 
Really, it is known starting from Ref.\cite{kel} that there is the 
possibility
to populate the ATI levels due to direct absorption of additional,
above-threshold quanta by the 
photoelectron. The probability of population of high ATI levels due to this
mechanism decreases rapidly, 
but it remains much higher, exponentially higher, than the probability of
absorption of the same number of quanta due to direct ionization 
of the second electron. 
The return of the photoelectron to the parent atomic particle and inelastic 
scattering are possible for any ATI level  excited due to
the direct mechanism Ref.\cite{k2}. It makes the scattering mechanism
productive. 

There is the other possibility for the population
of high ATI levels. 
Consider the following scenario. First the photoionization takes place.
Then the scattering of the photoelectron results in the population of a
high level in the ATI spectrum. After that the second collision takes place 
resulting in the double ionization.
 Analyzing  the classical trajectory it is easy to show
that the possibility for the second collision depends on the
energy level in the ATI spectrum.
If this level is below $f^2/2$ then the second collision can take place.
Otherwise,  
the photoelectron leaves the atomic particle after the first collision
and the considered scenario gives no contribution to the
excitation of the second electron. Therefore this scenario works only in the
restricted energy range. 
The given consideration shows that the
inelastic collision is possible for $k < f$. According
to Eq.(\ref{maxE}) it means that the ionization energy is restricted
by $E^{(A^{++})}-E^{(A^+)}\le 2 f^2$. 

In order to demonstrate the validity of the presented results, consider
the double ionization of an He atom. In the recent
experiment 
Ref.{\cite{wal}
the 780 nm laser field with intensity from 0.15 to 5.0 PW/cm$^2$ was
used. Even for the lowest intensity the yield of the doubly charged
ions was very pronounced. For intensity 0.15 PW/cm$^2$ the field momentum
is $f =1.11$. Therefore 
the maximal wiggling energy, $f^2/2=17.0$ eV is well below
the lowest excitation energy of He$^+$, which is  40.8 eV.
Absorption of only 3 quanta above the single-electron ionization threshold
changes the situation drastically. 
It gives the translational
momentum $k=0.642$. As a result the 
maximal energy of the  photoelectron, $(k+f)^2/2=42.2$ eV, becomes higher
than the excitation energy
allowing the excitation of He$^{+}$. 
Similarly, absorption of 6 - 7 quanta above the atomic threshold 
makes the  double ionization possible. Fig.1 shows the
terms of (e+He$^+$) and (e$_1$+e$_2$+He$^{++}$) systems given
by Eqs.(\ref{eA+}),(\ref{EA++}). Absorption of
7 quanta above the atomic 
threshold permits  crossing of the term {\bf b} of 
(e+He$^+$) with the terms {\bf d},{\bf e} 
of (e$_1$+e$_2$+He$^{++}$). Thus the latter
 can be populated and  double ionization takes place.
The probability of absorbing a few quanta above
the threshold is very 
high for the considered region of intensities of the laser 
field, see for example Ref.\cite{sch}.
Notice that 
there is the damping factor which describes how small is the  cross
section of an inelastic collision
in the laser field compared with the region of electron 
localization $\sigma_{\rm in}/R^2$. For estimation of the cross
section one can use the value $\sigma_{\rm in}\approx 
5\times 10^{-18}$cm$^2$, 
which according to Refs.\cite{in},\cite{exc}) describes the ionization
as well as the excitation field-free cross sections. If we estimate the
region of electron localization with the help of Eq.(\ref{R})
then for the intensity $0.15$ PW/cm$^2$ for 3 above-threshold quanta
we find $R=48$a$_0^2$, which gives  $\sigma_{\rm in}/R^2 \approx 
0.7\times10^{-4}$.
In Ref.\cite{wal} the same ratio was estimated using the modification of the
model Ref.\cite{cor}. This estimation gave the similar 
result $1.5\times 10^{-4}$.
In a recent work Ref.\cite{ku} there was  made an attempt 
to consider quantitatively the role of the rescattering mechanism
in the formation of the doubly charged ions of Helium. 
With this purpose the classical model was developed. 
The calculations based on this model give  the probability 
of the He$^{++}$
formation which is approximately by two orders of magnitude lower than the
experimental value of Ref.\cite{wal}. 

Notice that the rescattering mechanism in any case 
gives a huge enhancement compared with the sequential ionization. 
There is still a discrepancy between the numerical and
the experimental results. It was 
interpreted in Ref.\cite{ku} 
as an indication that the rescattering is not important for the considered
case,  and it was 
suggested that the violation of the adiabatic conditions during the 
removal of the first electron can be responsible for the effect.
There is a general physical argument against such a conclusion. 
In the initial wave function of Helium  the separation of the energy
levels greatly exceeds the frequency of the laser field. Therefore 
the adiabatic condition $\omega \ll \Delta E$ is well fulfilled. 
In this case the adiabatic theory predicts a very strong,
exponential suppression of  transitions from the ground state
into excited states, see Ref.\cite{la}.
 The situation drastically changes after
the ionization of the first electron, when  crossings
of the terms  become possible. 
At the moment of the crossing the adiabatic condition is
clearly  violated which results
in the population of the higher terms and provides the possibility
for  the double-charged ion  production.
Thus the rescattering 
is to be considered  as the preferred mechanism to explain 
the phenomena observed in Ref.\cite{wal}.
This statement seems to agree with the results of
calculations reported quite recently in Ref.\cite{bf}.

There are several points for uncertainties 
in all the calculated or estimated values mentioned above.
One of them is the poor knowledge of
the inelastic cross sections in the laser field. 
The other one is the model nature of the developed approaches. The main
assumption of this work, as well as of Ref.\cite{k2}, is the neglect
of the Coulomb field. From the physical point of view it is reasonable to 
assume that the Coulomb attraction can only but enhance the probability
of two-electron processes.

After single electron
ionization of the atom the term of the system $(e+A^+)$ 
in the laser field can undergo a crossing
with some term of the system $(e_1+e_2+A^{++})$ (or $(e+A^{+*})$). 
As a result a transition
from the first term to the second
one is  allowed and the second term
is highly populated.
This means that two-electron processes are very probable. 
ATI plays a very important role in this mechanism due to two reasons. First,
it results in a 
rise of the maxima of the 
terms of the system of $(e+A^+)$ after absorption of only
a few quanta above the atomic ionization threshold. Second, the
ATI states above the threshold of doubly charged ion give the 
strong decrease
of minima of the terms of the doubly charged ion.

I thank V.V.Flambaum, G.F.Gribakin, W.King,
 and O.P.Sushkov for valuable comments, and 
B.W.Boreham and H.Hora for reference \cite{bor}. 
The financial support of  the Australian Research Council is acknowledged.

\newpage
\begin{figure}[h]
\input psfig
\psfig{file=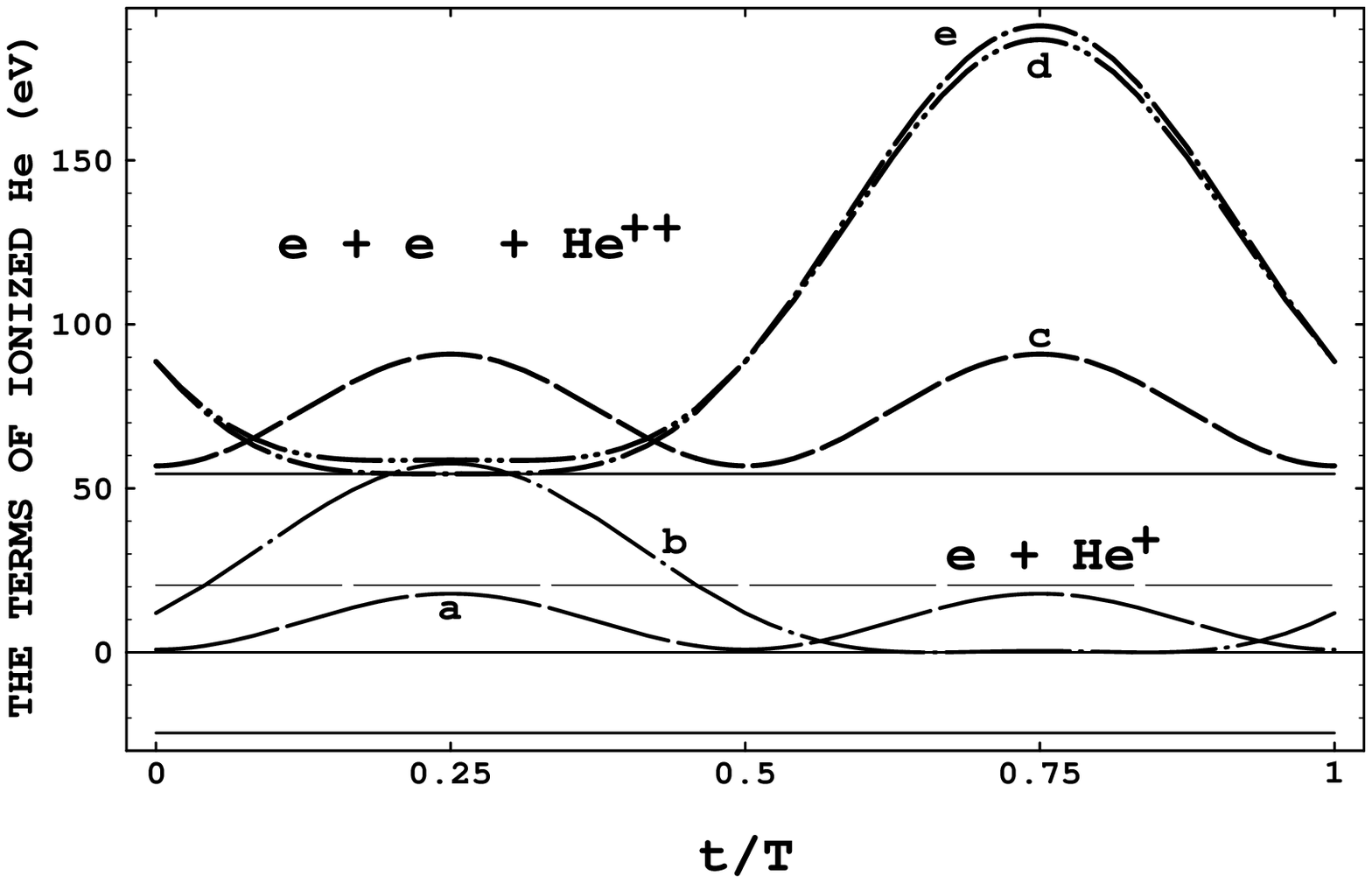, clip=}
\caption{\protect\small
Time evolution of the terms of the systems (e+He$^+$) and (e+e+He$^{++}$) 
given by Eqs.(3),(4).
$\omega = 2\pi/T=1.59$ eV corresponds to $\lambda=0.780$ nm.
The  field strength $F=0.0654~a.u$ corresponds to the lowest
intensity $1.5\times10^{14}$W/cm$^2$ in experiment Ref.[8].
The energy levels of He, He$^+$  and He$^{++}$
are shown by the horizontal solid lines.
{\bf a}.The ``unfavorable'' term of
(e+He$^+$): the lowest level above the atomic
threshold is occupied,
the available translational momentum is perpendicular to the field.
{\bf b}.The ``favorable'' 
term of (e+He$^+$): 7 above-threshold quanta are absorbed and
the translational momentum is parallel to the field.
The horizontal dashed
line is the position of this level in the ATI spectrum.
{\bf c}.The ``unfavorable'' term of (e+e+He$^{++}$): both electrons
occupy the lowest above-threshold level, their translational momenta
are perpendicular to the field.
{\bf d},{\bf e}.
The ``favorable'' terms of (e+e+He$^{++}$): 20 quanta above the 
threshold of He$^{++}$ are absorbed, the  translational electron momenta
are opposite to the field. 
{\bf d} and {\bf e} differ in the 
 the distribution of the above-threshold energy  E$_1$+E$_2$=
20 $\omega$ between the two electrons. {\bf d}.E$_1$  = E$_2=10\omega$. 
{\bf e}. 
E$_1=3\omega$, E$_2=17\omega$. Absorption of 7 quanta above the He 
threshold 
gives the crossing of the lower term {\bf b} with the upper
terms {\bf d} and {\bf e} resulting in double ionization.}
\end{figure}

\end{document}